\documentclass[journal]{IEEEtran}
\usepackage{amsfonts}
\usepackage{amssymb}
\usepackage{amsthm}
\usepackage{amsmath,amsfonts,amssymb}
\usepackage[dvips]{graphicx}
\usepackage{verbatim}
\usepackage{setspace}
\usepackage{bm}
\usepackage{subfigure} 
\usepackage{extarrows}
\usepackage{algorithmic} 
\usepackage[ruled,linesnumbered,vlined]{algorithm2e}
\usepackage{cite}

\usepackage{changepage}
\usepackage{pdfpages}
\usepackage{color}
\usepackage{bbding}

\newcommand{\figwidth}{8}
\IEEEoverridecommandlockouts
\usepackage{etoolbox}
\let\mybibitem\bibitem
\renewcommand{\bibitem}[1]{%
	\ifstrequal{#1}{nature}
	{\color{blue}\mybibitem{#1}}
	{\color{black}\mybibitem{#1}}%
}

\begin{document}

\title{Multiuser Communications with Movable-Antenna Base Station: Joint Antenna Positioning, Receive Combining, and Power Control}

\author{Zhenyu Xiao,~\IEEEmembership{Senior Member,~IEEE,}
	Xiangyu Pi,~\IEEEmembership{Graduate Student Member,~IEEE,}
	Lipeng Zhu,~\IEEEmembership{Member,~IEEE,}\\
	Xiang-Gen Xia,~\IEEEmembership{Fellow,~IEEE}
	and Rui Zhang,~\IEEEmembership{Fellow,~IEEE}
\thanks{Part of this paper has been submitted to IEEE Global Communications Conference 2023 Workshops~\cite{Conf2023Pi}.}
\thanks{Z.~Xiao and X.~Pi are with the School of Electronic and Information Engineering, Beihang University, Beijing 100191, China. (e-mail: xiaozy@buaa.edu.cn, pixiangyu@buaa.edu.cn).}
\thanks{L.~Zhu is with the Department of Electrical and Computer Engineering, National University of Singapore, Singapore 117583, Singapore. (e-mail: zhulp@nus.edu.sg).}
\thanks{X.-G. Xia is with the Department of Electrical and Computer Engineering, University of Delaware, Newark, DE 19716, USA. (e-mail: xxia@ee.udel.edu).}
\thanks{R. Zhang is with School of Science and Engineering, Shenzhen Research Institute of Big Data, The Chinese University of
	Hong Kong, Shenzhen, Guangdong 518172, China (e-mail:rzhang@cuhk.edu.cn). He is also with the Department of Electrical
	and Computer Engineering, National University of Singapore, Singapore 117583 (e-mail: elezhang@nus.edu.sg).}
}

\maketitle

\begin{abstract}
Movable antenna (MA) is an emerging technology which enables a local movement of the antenna in the transmitter/receiver region for improving the channel condition and communication performance. 
In this paper, we study the deployment of multiple MAs at the base station (BS) for enhancing the multiuser communication performance. 
First, we model the multiuser channel in the uplink to characterize the wireless channel variation due to MAs' movements at the BS. Then, an optimization problem is formulated to maximize the minimum achievable rate among multiple users for MA-aided uplink multiuser communications by jointly optimizing the MAs' positions, their receive combining at the BS, and the transmit power of users, under the constraints of finite moving region for MAs, minimum inter-MA distance, and maximum transmit power of each user. 
To solve this challenging non-convex optimization problem, a two-loop iterative algorithm is proposed by leveraging the particle swarm optimization (PSO) method. Specifically, the outer-loop updates the positions of a set of particles, where each particle's position represents one realization of the antenna position vector (APV) of all MAs. The inner-loop implements the fitness evaluation for each particle in terms of the max-min achievable rate of multiple users with its corresponding APV, where the receive combining matrix of the BS and the transmit power of each user are optimized by applying the block coordinate descent (BCD) technique. Simulation results show that the antenna position optimization for MAs-aided BSs can significantly improve the rate performance as compared to conventional BSs with fixed-position antennas (FPAs). 
\end{abstract}

\begin{IEEEkeywords}
Movable antenna (MA), antenna positioning, uplink communication, particle swarm optimization (PSO).
\end{IEEEkeywords}

\section{Introduction}
\IEEEPARstart{W}{ith} the coming of sixth-generation (6G) and beyond wireless communication systems, there is an urgent need for exploring large-capacity and high-reliability communication technologies~\cite{6G2020SW,Ma20216G,Wang20236G}. To achieve this goal, multi-user/multi-antenna or so-called multiple-input multiple-output (MIMO) communication technologies have been widely investigated to improve the spectral and energy efficiencies by exploiting the spatial multiplexing~\cite{SM2011RM,HBF2017MA,MIMO2021LQ}. Moreover, with the exploration of wireless communication systems migrating to higher frequency bands, such as millimeter-wave (mmWave) and terahertz (THz) bands, the smaller wavelength facilitates the implementation of large-scale MIMO systems, i.e, massive MIMO, to obtain higher beamforming and spatial multiplexing gains~\cite{MAMIMO2014LE,mmWave2014SSA,Zhu2019MIMO,Ning2020MIMO}. In addition, the explosive communication traffic growth in hotspot areas renders the multiuser MIMO (MU-MIMO) communication a crucial technique to meet the high throughput requirement~\cite{SDM2014AA,MUMIMO2018LF,MUMIMO2021DC}.

However, the antennas in conventional MIMO systems are deployed at fixed positions,  which cannot fully exploit the degrees of freedom (DoFs) in the continuous spatial domain for optimizing the spatial diversity/multiplexing performance. In order to overcome this fundamental limitation, movable antenna (MA) has been recently proposed as a new solution for fully exploiting the wireless channel variation in the continuous spatial domain~\cite{zhu2023movable,zhu2022modeling,ma2022mimo,MUMA2023ZL}. Different from conventional fixed-position antennas (FPAs), each MA is connected to a radio frequency (RF) chain via a flexible cable, which allows its position to be flexibly adjusted in a given spatial region with the aid of a driver component or by other means, for achieving more favorable channels to enhance the communication performance.
Meanwhile, to efficiently suppress the multiuser interference in the MU-MIMO system, each MA deployed in the transmitter/receiver region can move from a position experiencing severe interference to another position with weak interference. Consequently, the spatial DoFs can be fully exploited by independently adjusting the positions of different MAs, such that the spatial diversity and multiplexing performance of MA-aided communication systems is significantly improved. Although the spatial DoFs can also be achieved by conventional MIMO systems with antenna selection, they in general require a large number of antennas to overspead entire spatial regions for fully exploiting the spatial diversity and multiplexing gains~\cite{AS2004}. In comparison, such DoFs can be fully exploited by smaller number of MAs (or a single MA) deployed at the transmitter/receiver~\cite{zhu2023movable,zhu2022modeling}.

Prior studies have validated the superior communication performance of MA-aided systems to conventional FPA-aided systems. 
In~\cite{zhu2023movable}, the hardware architecture and channel characterization for MA systems were presented, and the advantages of MAs over conventional FPAs were demonstrated in terms of signal power improvement,
interference mitigation, flexible beamforming, and spatial multiplexing gain.
In~\cite{zhu2022modeling}, a field-response based channel model for MA-aided communication systems was developed, which characterized the channel variation with respect to MAs' positions. In addition, the conditions were derived under which the field-response based channel model is degraded to the well-known line-of-sight (LoS) channel, geometric channel, Rayleigh and Rician fading channel models. Moreover, under both deterministic and stochastic channel models, the maximum channel gain achieved by a single receive MA was analyzed, which demonstrated that the MA system can obtain considerable performance gains over FPA systems. 
Under the field-response based channel model, the channel capacity of the MA-aided MIMO system was maximized in~\cite{ma2022mimo} by jointly adjusting multiple MAs' positions in the transmitter and receiver located regions. Moreover, it was validated that jointly designing the positions of transmit and receive MAs can improve the spatial multiplexing performance of MIMO systems.
Besides, the total transmit power of multiple users was minimized in~\cite{MUMA2023ZL} by jointly optimizing the single-MA position and the transmit power of users, as well as the receive combining matrix of FPAs at the base station (BS) under the uplink multiuser communication setting. It was demonstrated that the MA-aided multiuser system can not only increase the channel gain, but also achieve more effective interference mitigation than that with FPAs.
In~\cite{chen2023joint}, the ergodic achievable rate for point-to-point MA-enhanced MIMO systems was maximized by jointly optimizing the MA positions and the transmit covariance matrix based on statistical channel state information (CSI) between the transmitter and receiver. 
In~\cite{wu2023movable}, the total transmit power of the BS was minimized for an
MA-enabled downlink multiuser communication system by optimizing the beamforming and discrete antenna positioning at the BS while guaranteeing
the minimum signal-to-interference-plus-noise ratio (SINR) of each user.
Note that the above performance improvement of MA systems relies on the availability of complete CSI between the entire transmit and receive regions where the MAs are located. 
To this end, a  novel successive transmitter-receiver compressed sensing (STRCS) method was proposed in~\cite{ma2023compressed} for channel estimation in  MA-aided communication systems, where the complete CSI between the transmit and receive regions was reconstructed by estimating the channel multi-path components based on the channel measurements taken at only a sufficient number of MAs' positions.
In addition, the authors in \cite{FAS2021Wk} proposed a novel fluid antenna system (FAS), which can be regarded as an alternative way for implementing MAs. Specifically, the physical position of an antenna can be switched to one of the predefined ports over a fixed-length line space, for selecting the position with the strongest channel to improve the communication performance. Furthermore, a fluid antenna multiple access (FAMA) system was proposed in~\cite{FAMA2022WK}, by controlling the positions of the fluid antennas to eliminate interference in multiuser communication systems.

However, none of the existing works~\cite{zhu2023movable,zhu2022modeling,ma2022mimo,MUMA2023ZL,ma2023compressed,chen2023joint, wu2023movable,FAS2021Wk,FAMA2022WK} investigated the employment of continuous-position MAs at the BS  to enhance the uplink communication performance of multiple users. Thus, the problem of exploiting multiple MAs at the BS side to serve multiple users remains unaddressed.  
In addition, to guarantee the performance fairness among multiple users in the  uplink communication, it is necessary to properly design the transmit power of each user and the multi-antenna receive combining at the BS.
Although there have been rich prior related works on receive combining at the BS and transmit power control at the users for multiuser uplink communication systems
\cite{PandW2000A,PandW2010J,PandW2012W}, they considered conventional FPA-based systems and their results cannot be directly applied for the MA-aided systems with controllable channels by antenna position optimization. As such, in this paper, we investigate a new MA-aided uplink multiuser communication system where multiple MAs are deployed at the BS to serve multiple users simultaneously, and each user is equipped with a single FPA. In particular, we study the joint optimization of MAs positioning, receive combining at the BS, and transmit power of different users to maximize their minimum achievable rate. 
The main contributions of this paper are summarized as follows: 
\begin{enumerate}
	\item We propose to deploy multiple MAs at the BS to improve the performance of a multiuser uplink communication system. This multiple-access channel (MAC) is modeled as a function of the antenna position vector (APV) to characterize the multi-path response between the multiple MAs at the BS and the single FPA at each user. Based on this channel model, we formulate an optimization problem to maximize the minimum achievable rate among multiple users by jointly designing the MAs' positions, their receive combining matrix at the BS and transmit power of each user, subject to the constraints of finite moving region for MAs, minimum inter-MA distance, and maximum transmit power of each user.
	
	\item To solve the formulated non-convex optimization problem with highly-coupled variables, a two-loop iterative algorithm is developed based on particle swarm optimization (PSO) to obtain a suboptimal solution efficiently. In the inner-loop, for a given APV, transmit power of each user and receive combining matrix at the BS are jointly designed by applying the block coordinate descent (BCD) technique, where the max-min achievable rate of multiple users is obtained by bisection search. In the outer-loop, a PSO-based algorithm is proposed to optimize the APV, where the fitness function of each particle (i.e., APV) is taken as the max-min achievable rate obtained in the inner-loop.
	
	\item Extensive simulations are conducted to evaluate the communication performance of our proposed solution for MA-aided multiuser communication systems. It is shown that the proposed scheme not only increases the channel power gain of multiple users but also suppresses the interference among them. Moreover, the simulation results 
	demonstrate that compared to conventional BS with FPAs, the proposed new BS architecture with MAs can significantly improve the rate performance by antenna position optimization.
	Besides, we evaluate the impact of imperfect field-response information (FRI) on the optimized solution for APV. The results reveal that although the imperfectly estimated angles and coefficients of the channel paths lead to certain performance degradation, the proposed solution can still obtain considerable performance gain over FPA-aided BS.
\end{enumerate} 

Note that compared to the prior work~\cite{MUMA2023ZL}, we consider a different  scheme in which the multiple MAs are deployed at the BS side instead of each at the user side. In general, there are more space and energy available at the BS side, which makes MAs' deployment and movement more practically feasible. As for the optimization problem, we focus on maximizing the minimum achievable rate among all users to ensure fairness, while \cite{MUMA2023ZL} aims to minimize the total transmit power of users given their rate requirements. In addition, since each user is equipped with a single MA in~\cite{MUMA2023ZL}, its movement only changes the channel between its corresponding user and the BS. In contrast, in our considered system, all MAs are deployed in the same receive region at the BS and the movement of each MA results in a change of all users' channels in general.

The rest of this paper is organized as follows. In Section~\ref{sec_SystemModel}, we introduce the system model of the considered MA-aided multiuser communication system and formulate the 
optimization problem. In Section~\ref{sec_ProposedSolution}, we introduce the proposed solution for the
formulated optimization problem and discuss its convergence and computational complexities. 
Section~\ref{sec_SimulationResults} presents the simulation results. Finally, the paper is concluded in Section~\ref{sec_Conclusion}.

\textit{Notation}: $a$, $\mathbf{a}$, $\mathbf{A}$, and $\mathcal{A}$ denote a scalar, a vector, a matrix, and a set, respectively. $\mathbb R$ and $\mathbb C$ represent the sets of real and complex numbers, respectively. $\mathbb{R}^M$ and $\mathbb{C}^M$ denote the space of $M$-dimensional real vectors and complex vectors, respectively. $(\cdot)^{\rm{T}}$, $(\cdot)^{*}$, and $(\cdot)^{\rm{H}}$ denote transpose, conjugate, and conjugate transpose, respectively. $|\mathcal{A}|$ denotes the cardinality of set $\mathcal{A}$. $[\mathbf{a}]_i$ and $[\mathbf{A}]_{i,j}$ denote the $i$-th entry of vector $\mathbf{a}$ and the entry in the $i$-th row and $j$-th column of
matrix $\mathbf{A}$, respectively.
$\mathrm{Max}\{{\mathbf{a}}\}$ and $\mathrm{Min}\{{\mathbf{a}}\}$ are the maximum and minimum entry of real vector $\mathbf{a}$, respectively.
 $\|\mathbf{a}\|_2$ represents the L2-norm of vector $\mathbf{a}$. 
$\mathrm{diag}\{\mathbf{a}\}$ is a diagonal matrix
with the entry in the $i$-th row and $i$-th column equal to the $i$-th entry of vector $\mathbf{a}$. $\mathbf{I}_N$ denotes the identity matrix of size $N\times N$. $\mathcal{CN}(0,\mathbf{A})$ represents the circularly symmetric
complex Gaussian (CSCG) distribution with mean zero and covariance matrix $\mathbf{A}$. $\mathcal{U}[a,b]$ denotes the uniform distribution over the real-number interval $[a,b]$. 

\section{System Model and Problem Formulation}\label{sec_SystemModel}

\begin{figure}[!t]
	\centering
	\includegraphics[width=8.8 cm]{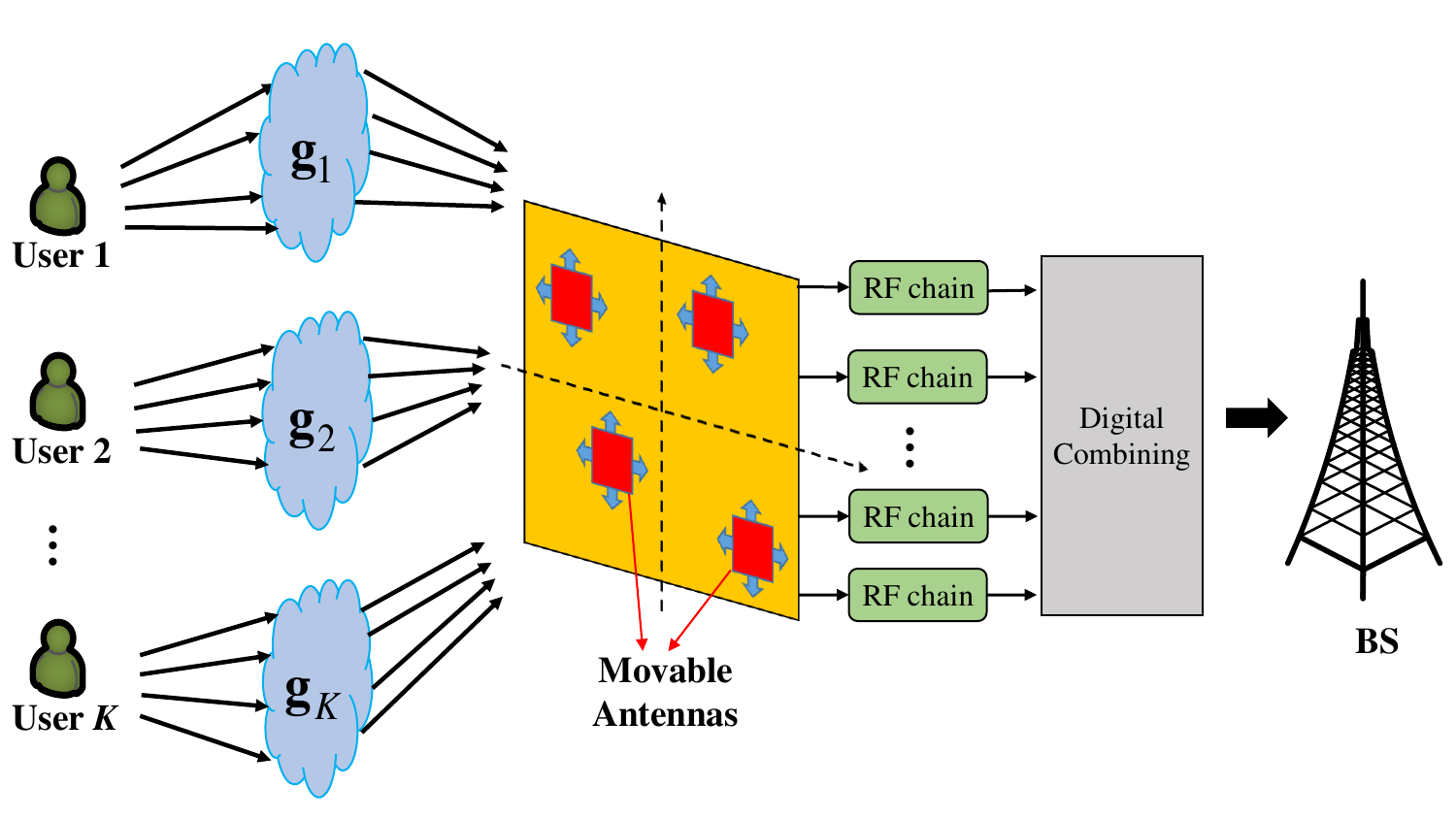}
	\caption{The MA-aided multiuser communication system.}
	\label{fig:system_model}
\end{figure}
As shown in Fig.~\ref{fig:system_model}, $K$ single-FPA users are served by the BS equipped with $M$ MAs, each of which is connected to an RF chain via a flexible cable, thus enabled to move in a local two-dimensional (2D) region $\mathcal{C}_{r}$ at the BS for improving the channel conditions with users. We consider the space-division multiple access (SDMA) of users simultaneously communicating with the BS in the uplink at a given frequency band, and thus the number of users is assumed not exceeding that of MAs at the BS, i.e., $K\le M$. The position of the $m$-th receive MA can be represented by its Cartesian coordinates, $\mathbf{r}_m=[x_m,y_m]^\mathrm{T} \in \mathcal{C}_{r}$ for $  1 \leq m \leq M$. Without loss of generality, the 2D region for antenna moving, i.e., $ \mathcal{C}_{r}$, is assumed as a square region of size $A\times A$. 

The received signal at the BS is processed using a digital combining matrix,
\begin{equation}
	\label{signal_model}
	\mathbf{y}=\mathbf{W}^{\mathrm{H}} \mathbf{H}(\tilde{\mathbf{r}}) \mathbf{P}^{1 / 2} \mathbf{s}+\mathbf{W}^{\mathrm{H}} \mathbf{n},
\end{equation}
where $\mathbf{W} =\left[\mathbf w_1,\mathbf w_2,\cdots,\mathbf w_K\right]\in\mathbb C^{M\times K}$  is the receive combining matrix at the BS, $\mathbf{H}(\tilde{\mathbf{r}})=[\mathbf{h}_1(\tilde{\mathbf{r}}),\mathbf{h}_2(\tilde{\mathbf{r}}),\cdots,\mathbf{h}_K(\tilde{\mathbf{r}})]\in\mathbb{C}^{M\times K}$ is the channel matrix from all $K$ users to the $M$ MAs at the BS with $\tilde{\mathbf{r}}=[\mathbf{r}^\mathrm{T}_1,\mathbf{r}^\mathrm{T}_2,\cdots,\mathbf{r}^\mathrm{T}_M]^\mathrm{T}$ denoting the APV for MAs, $\mathbf{P}^{1/2}=\mathrm{diag}\left\{\sqrt p_1,\sqrt p_2,\cdots,\sqrt {p_K}\right\}$ is the power matrix with $p_k , 1 \leq k \leq K$, representing the transmit power of user $k$, $\mathbf{s}$ is the independent
and identically distributed (i.i.d.) transmit signal vector of users each with normalized power, i.e., $\mathbb{E}(\mathbf{s}\mathbf{s}^{\mathrm{H}})=\mathbf{I}_K$, and $\mathbf{n}\sim \mathcal{CN}(0,\sigma^2\mathbf{I}_M)$ is the zero-mean additive white Gaussian noise (AWGN) with vector with covariance matrix $\sigma^2\mathbf{I}_M$.
\subsection{Channel Model}
We employ the field-response based channel model in~\cite{zhu2022modeling}, where the channel response is the superposition of the coefficients of multiple channel paths between the transceivers. For the considered MA-aided multiuser communication system, we assume that the far-field condition is satisfied between the BS and users since the size of the moving region for the MAs at the BS is much smaller than the signal propagation distance. Thus, for each user, the angles of arrival (AoAs) and the amplitudes of the complex path coefficients for multiple channel paths do not change for different positions of the MAs, which means that only the phases of the multiple channel paths vary in the receive region.

Let $L_{k}$ denote the total number of receive channel paths at the BS from user $k$, $1 \leq k \leq K$. Then, the signal propagation phase difference of the $l$-th path  for user $k$ between the position of the $m$-th MA and the reference point at the BS, $\mathbf{r}_{0}=[0, 0]^T$, is written as 
\begin{equation}
	\label{phase difference}
	\rho_{k,l}(\mathbf{r}_m)=x_m \sin{\theta_{k,l}}\cos{\phi_{k,l}}+y_m \cos{\theta_{k,l}},
\end{equation}
where $\theta_{k,l}$ and $\phi_{k,l}$  are the elevation and azimuth AoAs for the $l$-th receive path between the user $k$ and the BS. Accordingly, the field-response vector (FRV) of the receive channel paths between the user $k$ and the $m$-th MA at the BS is given by ~\cite{zhu2022modeling}
\begin{equation}
	\label{FRV}
	\mathbf{f}_k(\mathbf{r}_m)=\left[e^{j\frac{2\pi}{\lambda}\rho_{k,1}(\mathbf{r}_m)},e^{j\frac{2\pi}{\lambda}\rho_{k,2}(\mathbf{r}_m)},\dots,e^{j\frac{2\pi}{\lambda}\rho_{k,L_k}(\mathbf{r}_m)}\right]^\mathrm{T}.
\end{equation}
As such, the channel vector between user $k$ and the BS is obtained as 
\begin{equation}
	\label{channel_model}
	\mathbf{h}_k(\tilde{\mathbf{r}})=\mathbf{F}_k^{\mathrm{H}}(\tilde{\mathbf{r}})\mathbf{g}_k,
\end{equation}
where $\mathbf{F}_k(\tilde{\mathbf{r}})=[\mathbf{f}_k(\mathbf{r}_1),\mathbf{f}_k(\mathbf{r}_2),\cdots,\mathbf{f}_k(\mathbf{r}_M)]\in\mathbb{C}^{L_k\times M}$ denotes the field-response matrix (FRM) at the BS,
and $\mathbf{g}_k=[g_{k,1},g_{k,2},\cdots,g_{k,L_k}]^{\mathrm T}$ is the path-response vector (PRV), which represents the multi-path response coefficients from user $k$ to the reference point in the receive region. As can be observed, the channel coefficient $[\mathbf{h}_k(\tilde{\mathbf{r}})]_m=\mathbf{f}_k(\mathbf{r}_m)^{\mathrm H}\mathbf{g}_k$ between the $m$-th MA and user $k$ is the sum of all elements of $\mathbf{g}_k$ weighted by the unit-modulus elements in $\mathbf{f}_k(\mathbf{r}_m)^{\mathrm H}$.  
As a result, small movement of each MA can change the channel vectors of all users significantly due to the phase variations of multiple channel paths (while their amplitude variations are relatively much less and thus negligible).
\subsection{Problem Formulation}
At the BS, the receive SINR for user $k$ is given by
\begin{equation}
	\label{SINR}
	\gamma_k=\frac{\left|\mathbf{w}_k^{\mathrm{H}}\mathbf{h}_k(\tilde{\mathbf{r}})\right|^2p_k}
	{\sum\limits _{i=1,i\ne k}^{K}\left|\mathbf{w}^{\mathrm{H}}_k\mathbf{h}_i(\tilde{\mathbf{r}})\right|^2p_i+\left\|\mathbf{w}_{k}\right\|_2^2\sigma^2}.
\end{equation}
Thus, the achievable rate for user $k$ is calculated as
\begin{equation}
	\label{Rate}
	R_k=\log_{2}{\left (1+ \gamma_k\right)}.
\end{equation}
It should be emphasized that different from conventional FPAs, the achievable rate of each user depends on the APV $\tilde{\mathbf{r}}$, which determines the channel matrix $\mathbf{H}(\tilde{\mathbf{r}})$ and thus influences the corresponding optimal receive combining matrix $\mathbf{W}$ as well as the transmit power matrix $\mathbf{P}$. 

In this paper, we aim to maximize the minimum achievable rate among all users to improve the overall performance by jointly optimizing the APV of MAs at the BS, i.e., $\tilde{\mathbf{r}}$, their receive combining matrix, i.e., $\mathbf{W}$, and the transmit power matrix, i.e., $\mathbf{P}$.
The max-min rate optimization problem is formulated as\footnote{The FRI in the angular domain, including AoAs and PRVs, is assumed to be known, which can be acquired by using channel estimation methods for MA systems, such as STRCS in~\cite{ma2023compressed}.
}
\begin{subequations}\label{op}
	\begin{align}
		\max \limits _{\tilde{\mathbf{r}},\mathbf W,\mathbf P}&~~\min \limits _{k}~~\{R_k\} \label{opA}\\
		\mbox { s.t.}~~ 
		& {\mathbf{r}}_m \in \mathcal{C}_{r}, 1 \le  m  \le M, \label{opB}\\
		& \left\|\mathbf{r}_{m}-\mathbf{r}_{i}\right\|_{2} \geq D, 1 \le  m \neq i \le M, \label{opC}\\
		& 0\le p_k \le p_{\mathrm{max}}, 1 \le k \le K.\label{opD}
	\end{align}   
\end{subequations}
Constraint (\ref{opB}) indicates that each MA can only move in the given receive region, $\mathcal{C}_{r}$. 
Constraint (\ref{opC}) ensures that minimum inter-MA distance $D$ at the BS for practical implementation. 
Constraint (\ref{opD}) ensures that transmit power of each user is non-negative and does
not exceed its maximum value, $p_{\mathrm{max}}$.

Note that problem (\ref{op}) is highly non-convex and challenging to solve because the objective function (\ref{opA}) is non-concave/non-convex over the APV $\tilde{\mathbf{r}}$, the receive combining matrix $\mathbf{W}$ and the transmit power matrix $\mathbf{P}$. Besides, the three high-dimensional vector/matrix variables are highly coupled with each other.
Existing optimization tools cannot be directly used to obtain the globally optimal solution for problem (\ref{op}) with polynomial complexity in terms of $M$ and $K$. Next, we will develop a two-loop iterative algorithm based on PSO to obtain a suboptimal solution for problem (\ref{op}).

\section{Proposed Solution}\label{sec_ProposedSolution}
\begin{figure}[!t]
	\centering
	\includegraphics[width=\figwidth cm]{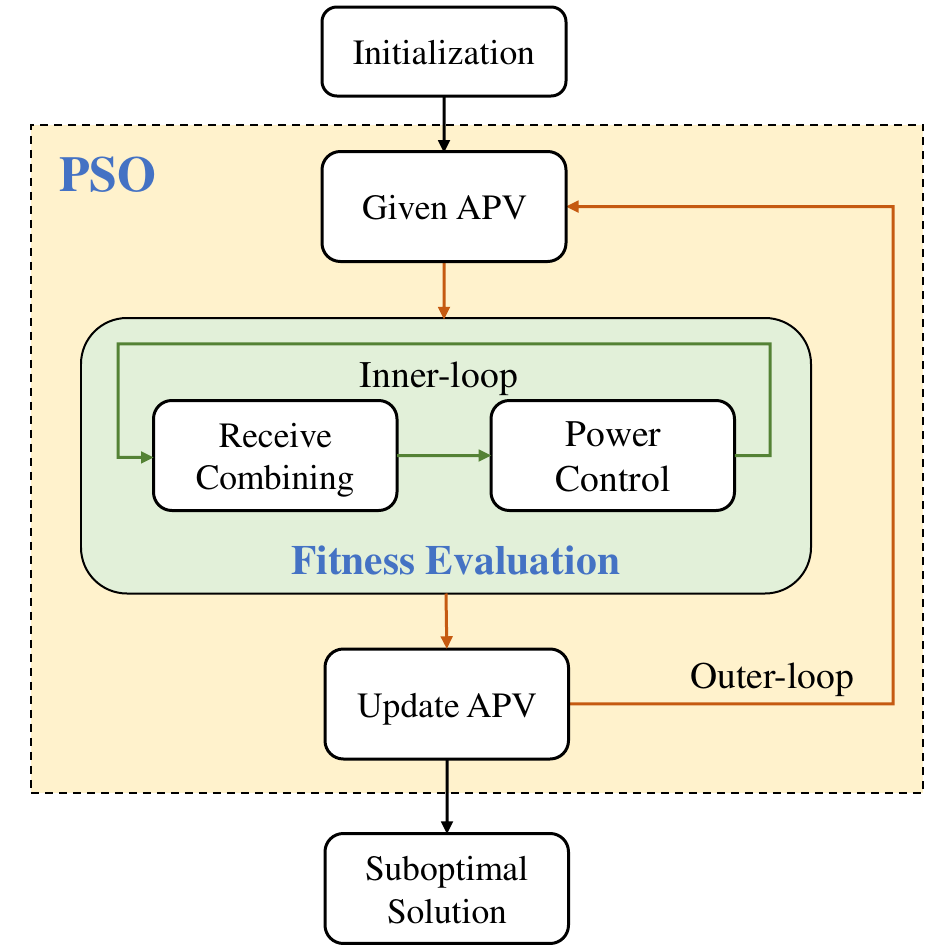}
	\caption{The flowchart of two-loop iterative algorithm for solving problem (\ref{op}).}
	\label{fig:flowchart}
\end{figure}
Since there are three highly coupled matrices/vectors in the optimization variables of problem (\ref{op}), the conventional alternating optimization method approach by optimizing one of them with the other two being fixed  may not work well as it may lead to an undesired local optimal solution. For example, the optimal receive combing matrix based on the given APV and transmit power matrix will narrow the optimization space of the APV to a tiny region around the given APV in the next iteration, since the channel vectors of other APV solutions do not match the receive combining matrix.
To address this problem, we propose a two-loop iterative algorithm based on PSO. In the inner-loop, for any given APV, a BCD-based algorithm is developed to iteratively solve the receive combining and transmit power optimization. In the outer-loop, a PSO-based algorithm is applied to optimize the APV, where the fitness function of each particle (i.e., APV) is the max-min achievable rate obtained in the inner-loop. The whole flowchart of the proposed two-loop iterative algorithm is shown in Fig.~\ref{fig:flowchart}.
\subsection{Receive Combining and Transmit Power Optimization}
In the inner-loop of the proposed algorithm, in order to calculate the fitness value of each particle, which represents an APV solution, we need to solve the following problem to determine the receive combining matrix and transmit power matrix for any given APV: 
\begin{subequations}\label{op:PandW}
	\begin{align}
		\max \limits _{\mathbf W,\mathbf P}&~~\min \limits _{k}~~\left\{R_k\right\} \label{op:PandW_A}\\
		\mbox { s.t.}~~
		&0\le p_k \le p_{\mathrm{max}}, 1 \le k \le K.\label{op:PandW_B}
	\end{align}
\end{subequations}
Problem (\ref{op:PandW}) have been previously investigated in existing literature~\cite{PandW2000A,PandW2010J,PandW2012W}, which, however, requires a high computational complexity in general. Since the solution for problem (\ref{op:PandW}) should be substituted to the outer-loop during the iterations, a low-complexity algorithm for solving problem (\ref{op:PandW}) is required. To this end, we develop a BCD-based algorithm with low computational complexity, in which the receive combining matrix and transmit power matrix are alternatively optimized with the other one being fixed.

Note that for any given APV $\tilde{\mathbf r}$ and transmit power matrix $\mathbf P$, 
the optimal receive combining matrix $\mathbf W$ can be derived in closed form based on the minimum mean square error (MMSE) receiver~\cite{MMSE2017Am,MMSE2020HH,MMSE2022Lin}, i.e.,
\begin{equation}
	\label{MMSE}
	\begin{split}
		\hat{\mathbf W}(\tilde{\mathbf r},\mathbf P)&=\left(\mathbf H(\tilde{\mathbf r})\mathbf P\mathbf{H}(\tilde{\mathbf r})^{\mathrm H}+\sigma^2\mathbf I_M\right)^{-1}\mathbf {H}(\tilde{\mathbf r})\\
		&\triangleq\left[\hat{\mathbf{w}}_1,\hat{\mathbf{w}}_2,\cdots,\hat{\mathbf{w}}_K\right],\\   
	\end{split}
\end{equation}
with $\hat{\mathbf{w}}_k=\left(\mathbf H(\tilde{\mathbf r})\mathbf P\mathbf{H}(\tilde{\mathbf r})^{\mathrm H}+\sigma^2\mathbf I_M\right)^{-1}\mathbf{h}_k(\tilde{\mathbf r})$. Substituting (\ref{MMSE}) into (\ref{SINR}), the receive SINR for user $k$ given in (\ref{SINR}) can be rewritten as
\begin{equation}
	\label{SINR_OPpower}
	\hat{\gamma}_k=\frac{p_k[\mathbf{A}]_{k,k}}
	{\sum\limits _{i=1,i\ne k}^{K}p_i[\mathbf{A}]_{k,i}+b_k},
\end{equation}
where $[\mathbf{A}]_{k,i}\triangleq \left|\hat{\mathbf{w}}^{\mathrm{H}}_k\mathbf{h}_i(\tilde{\mathbf{r}})\right|^2
= \left|\mathbf{h}^{\mathrm{H}}_k(\tilde{\mathbf r})\left(\mathbf H(\tilde{\mathbf r})\mathbf P\mathbf{H}(\tilde{\mathbf r})^{\mathrm H}+\sigma^2\mathbf I_M\right)^{-1}\mathbf{h}_i(\tilde{\mathbf{r}})\right|^2, 1 \leq k,i \leq K, $ is the entry in the $k$-th row and $i$-th column of matrix $\mathbf{A}\in\mathbb{C}^{K\times K}$ and $b_k \triangleq\left\|\hat{\mathbf{w}}_{k}\right\|_2^2\sigma^2
= \left\|\left(\mathbf H(\tilde{\mathbf r})\mathbf P\mathbf{H}(\tilde{\mathbf r})^{\mathrm H}+\sigma^2\mathbf I_M\right)^{-1}\mathbf{h}_k(\tilde{\mathbf r})\right\|_2^2\sigma^2, 1 \leq k \leq K, $ is the $k$-th entry of column vector $\mathbf{b}=[b_1,b_2,\cdots,b_K]^\mathrm{T}\in\mathbb{C}^{K\times 1}$.

For any given APV $\tilde{\mathbf r}$ and receive combining matrix $\mathbf W$, in order to distinguish with the transmit power matrix used to calculate the receive combining matrix in the previous iteration, we introduce the transmit power vector $\mathbf{p}=[p_1,p_2,\cdots,p_k]^{\mathrm{T}}$ as an intermediate variable in the current iteration.
Therefore,
problem (\ref{op:PandW}) can be equivalently transformed into
\begin{subequations}\label{op_power}
	\begin{align}
		\max \limits _{\mathbf p,\eta}&~~\eta \label{op_powerA}\\
		\mbox { s.t.}~~ 
		& \hat{\gamma}_k \ge \eta, 1 \le k \le K,\label{op_powerB}\\
		& 0\le p_k \le p_{\mathrm{max}}, 1 \le k \le K,\label{op_powerC}
	\end{align}   
\end{subequations} 
where $\eta$ represents the minimum SINR among the users. It is easy to verify that the optimal solution for problem (\ref{op_power}) is obtained as the constraints in (\ref{op_powerB}) are met with equality \cite{Maxmin2008Nace,Maxmin2021LYM}. Otherwise, we can always adjust the transmit power of certain users to ensure the equality holds with the minimum SINR unchanged.
In other words, the linear equations $p_k[\mathbf{A}]_{k,k}/\eta=\sum\limits _{i=1,i\ne k}^{K}p_i[\mathbf{A}]_{k,i}+b_k,1\le k \le K,$ always hold,
which is equivalent to the following matrix form of linear equations with respect to $\mathbf{p}$:
\begin{equation}
	\label{Linear_equation}
	\mathbf{D(\eta)} \mathbf{p} = \mathbf b,
\end{equation}
where $\mathbf{D(\eta)}\in\mathbb C^{K \times K}$ is a square matrix whose diagonal elements and non-diagonal elements are given by $[\mathbf{D}(\eta)]_{k,k}=[\mathbf{A}]_{k,k}/\eta$ and $[\mathbf{D}(\eta)]_{k,i}=-[\mathbf{A}]_{k,i}$ for $1 \le k \ne i \le K$, respectively.
Thus, the transmit power vector can be expressed as a function with respect to $\eta$:
\begin{equation}
	\label{power}
	{\mathbf{p}(\eta)} = \mathbf{D(\eta)}^{-1} \mathbf b.
\end{equation}

It is worth emphasizing that the solution for transmit power vector shown in  (\ref{power}) is feasible to problem (\ref{op_power}) only if constraint (\ref{op_powerC}) is satisfied. Thus, we develop the bisection method to find the maximum $\eta$ which makes ${\mathbf{p}(\eta)}$ satisfy constraint (\ref{op_powerC}). First, we choose an initial search interval $(\eta_{\mathrm{min}},\eta_{\mathrm{max}})$ 
that contains the optimal $\eta$. 
Then, the feasibility of the middle point of search internal,  $\eta=(\eta_{\mathrm{min}}+\eta_{\mathrm{max}})/{2}$, is examined by checking
whether $\mathrm{Min}\{\mathbf{p}(\eta)\} \ge 0$ and $\mathrm{Max}\{\mathbf p\}\le p_{\mathrm{max}}$. If $\eta$ is feasible, we update $\eta_{\mathrm{min}}$ as $\eta$, and otherwise update $\eta_{\mathrm{max}}$ as $\eta$. This process is repeated until the predefined accuracy is achieved.

The detailed bisection-based algorithm is shown in Algorithm~\ref{bisection}.
In line 1, we first initialize the lower bound and upper bound of $\eta$ with $\eta_{\mathrm{min}}=0$ and $\eta_{\mathrm{max}}=p_{\mathrm{max}}h_{\mathrm{min}}/\sigma^2$, where $h_{\mathrm{min}}$ is the minimum channel gain among users, i.e.,
$h_{\mathrm{min}}=\mathrm{Min}\left\{ \left\|\mathbf{h}_{1}(\tilde{\mathbf{r}})\right\|_2^2,\left\|\mathbf{h}_{2}(\tilde{\mathbf{r}})\right\|_2^2,\cdots,\left\|\mathbf{h}_{k}(\tilde{\mathbf{r}})\right\|_2^2\right\}$. 
In lines 3-4, we calculate the transmit power vector according to (\ref{Linear_equation}) for given $\eta=(\eta_{\mathrm{min}}+\eta_{\mathrm{max}})/2$. 
In lines 5-9, if the transmit power of each user satisfies constraint (\ref{op_powerC}),
then the lower bound $\eta_{\mathrm{min}}$ is replaced with the current $\eta$. Otherwise, the upper bound $\eta_{\mathrm{max}}$ is replaced with the current $\eta$. Therefore, in lines 2-10, the transit power vector updates until the search interval is less than the positive convergence threshold $\epsilon$.  
\begin{algorithm}[t] \small
	\label{bisection}
	\caption{Bisection-based algorithm for solving problem~(\ref{op_power}).}
	\begin{algorithmic}[1]
		\REQUIRE ~$\mathbf{H}(\tilde{\mathbf r})$, $\mathbf W$, $p_{\mathrm{max}}$, $\sigma^2$, $\epsilon$. 
		\ENSURE ~$\mathbf p$. \\
		\STATE Initialize $\eta_{\mathrm{min}}=0$ and $\eta_{\mathrm{max}}=\frac{p_{\mathrm{max}}h_{\mathrm{min}}}{\sigma^2}$ that contains the optimal value.
		\WHILE {$\eta_{\mathrm{max}} - \eta_{\mathrm{min}} >\epsilon$}
		\STATE Calculate the middle point $\eta=\frac{\eta_{\mathrm{min}}+\eta_{\mathrm{max}}}{2}$ of search\\ interval $(\eta_{\mathrm{min}},\eta_{\mathrm{max}})$.
		\STATE Calculate $\mathbf{p}(\eta)$ according to (\ref{power}) for given $\eta$.
		\IF  {$\mathrm{Min}\{\mathbf{p}(\eta)\} \ge 0$ and $\mathrm{Max}\{\mathbf{p}(\eta)\} \le p_{\mathrm{max}}$ }
		\STATE Update $\eta_{\mathrm{min}}\leftarrow \eta$.
		\ELSE
		\STATE Update $\eta_{\mathrm{max}} \leftarrow \eta$.	
		\ENDIF
		\ENDWHILE
		\RETURN ${\mathbf p}$.
	\end{algorithmic}
\end{algorithm}

Based on the above analysis, a BCD-based algorithm is developed to jointly optimize the receive combining matrix and transmit power matrix with given APV $\tilde{\mathbf r}$. In each iteration, for given transmit power matrix $\mathbf{P}$, we obtain a closed-form solution for receive combining matrix $\mathbf{W}$ according to (\ref{MMSE}). For given $\mathbf{W}$, we then solve the transmit power vector $\mathbf{p}$ by using Algorithm~\ref{bisection}, and update it into $\mathbf{P}=\mathrm{diag}\{\mathbf{p}\}$. 
During the iterations, the receive combining matrix and transmit power matrix are alternately optimized until convergence. 
The detailed BCD-based algorithm is shown in Algorithm~\ref{BCD}.
In line 1, the transmit power of all users is initialized to the maximum power, i.e., $\mathbf{P}^{(0)}=p_{\mathrm{max}}\mathbf{I}_K$. Then the channel matrix $\mathbf{H}(\tilde{\mathbf r})$ is calculated in line 2. With the input $\mathbf{P}^{(0)}$ and $\mathbf{H}(\tilde{\mathbf r})$, the initial receive combining matrix $\mathbf W^{(0)}$ is obtained by the MMSE receiver in line 3. Subsequently, the receive combining matrix and transmit power matrix are alternately optimized in lines 4-10 until convergence. Note that in line 8, the minimum achievable rate among multiple users in the $j$-th iteration is defined as
\begin{equation}
	\label{RATE_PandV}
	\mathcal{G}\left(\mathbf{P}^{(j)},\mathbf W^{(j)}\right) = \operatorname*{min}_{k}~\left\{R_k\right\},
\end{equation}
where $R_k$ can be calculated by (\ref{SINR}) and (\ref{Rate}) with given $\mathbf{P}^{(j)}$ and $\mathbf W^{(j)}$. The iteration process will terminate if the relative increase of the objective value is below a positive convergence threshold $\xi$. Finally, the optimal receive combining matrix and transmit power matrix are obtained, which correspond to the max-min achievable rate of multiple users for the given APV, i.e., $R(\tilde{\mathbf r}) =\mathcal{G}\left(\mathbf{P},\mathbf W\right)$.
\begin{algorithm}[t] \small
	\label{BCD}
	\caption{BCD-based algorithm for solving problem~(\ref{op:PandW}).}
	\begin{algorithmic}[1]
		\REQUIRE ~$\tilde{\mathbf r}$, $M$, $K$, $\lambda$, $p_{\mathrm{max}}$, $\sigma^2$, $\{g_k\}$, $\{\theta_{k,l}\}$, $\{\phi_{k,l}\}$, $\epsilon$, $\xi$.
		\ENSURE ~$\mathbf W$, $\mathbf P$, $R(\tilde{\mathbf r})$. \\
		\STATE Set the iteration index as $j=1$ and initialize
		$\mathbf{P}^{(0)}=p_{\mathrm{max}}\mathbf{I}_K$.  
		\STATE Calculate the channel response matrix $\mathbf{H}(\tilde{\mathbf r})$ according to (\ref{channel_model}) for given $\tilde{\mathbf r}$.
		\STATE Initialize the receive combining matrix $\mathbf W^{(0)}$ according to (\ref{MMSE}) for given $\mathbf{P}^{(0)}$ and $\tilde{\mathbf r}$.
		\REPEAT
		\STATE Calculate the transmit power vector $\mathbf{p}^{(j)}$ according to Algorithm \ref{bisection} with the input $\mathbf{W}^{(j-1)}$ and $\mathbf{H}(\tilde{\mathbf r})$.
		\STATE Update $\mathbf{P}^{(j)}=\text{diag}\{\mathbf{p}^{(j)}\}$.
		\STATE Calculate the receive combining matrix $\mathbf W^{(j)}$ according to (\ref{MMSE}) for given $\mathbf{P}^{(j)}$ and $\mathbf{H}(\tilde{\mathbf r})$.
		\STATE Calculate $\mathcal{G}\left(\mathbf{P}^{(j)},\mathbf W^{(j)}\right)=\operatorname*{min}\limits_{k}~\left\{R_k\right\}$ for given $\mathbf{P}^{(j)}$ \quad and $\mathbf{W}^{(j)}$.
		\STATE Update $j\leftarrow j+1$.
		\UNTIL{$\left|\mathcal{G}\left(\mathbf{P}^{(j)},\mathbf W^{(j)}\right)-\mathcal{G}\left(\mathbf{P}^{(j-1)},\mathbf W^{(j-1)}\right)\right|<\xi$}
		\STATE Set the transmit power matrix $\mathbf P$ as $\mathbf{P}^{(j)}$.
		\STATE Set the receive combining matrix $\mathbf W$ as $\mathbf W^{(j)}$ .
		\STATE Calculate the maximum objective value for given APV $\tilde{\mathbf r}$, $R(\tilde{\mathbf r}) =\mathcal{G}\left(\mathbf{P},\mathbf W\right)$.
		\RETURN  $\mathbf W$, $\mathbf P$, $R(\tilde{\mathbf r})$.
	\end{algorithmic}
\end{algorithm}


\subsection{APV Optimization}
In the outer-loop of proposed algorithm, since the optimal receive combining matrix and transmit power matrix for any given APV $\tilde{\mathbf{r}}$ can be calculated in the inner-loop, the corresponding max-min achievable rate for multiple users can be accordingly expressed as a function for the APV, i.e.,  $R(\tilde{\mathbf{r}})$. Thus, the original problem~(\ref{op}) can be transformed to the following APV optimization problem
\begin{subequations}\label{op_APV}
	\begin{align}
		\max \limits _{\tilde{\mathbf{r}}}&~~R(\tilde{\mathbf{r}}) \label{op_APVA}\\
		\mbox { s.t.}~~ 
		& {\mathbf{r}}_m \in \mathcal{C}_{r}, 1 \le  m  \le M \label{op_APVB}\\
		& \left\|\mathbf{r}_{m}-\mathbf{r}_{i}\right\|_{2} \geq D, 1 \le  m \neq i \le M. \label{op_APVC}
	\end{align}   
\end{subequations}
Although $R(\tilde{\mathbf{r}})$ can be calculated
according to Algorithm~\ref{BCD}, the highly non-convex
form makes it complicated to solve problem
(\ref{op_APV}) directly. In addition, the solution space of the APV, i.e.,
$[-A/2, A/2]^{2M}$, is large in general, so it may result in a computationally prohibitive complexity to
directly search for the optimal solution.
To solve this difficult problem, PSO is introduced as an efficient approach~\cite{PSO2009Yao,PSO2019Zhu,mao2023joint}.

In the PSO-based algorithm, we first randomly initialize $N$ particles with positions $\mathcal{R}^{(0)}=\{\tilde{\mathbf{r}}_1^{(0)},\tilde{\mathbf{r}}_2^{(0)},...,\tilde{\mathbf{r}}_{N}^{(0)}\}$  and velocities  $\mathcal{V}^{(0)}=\{\tilde{\mathbf{v}}_1^{(0)},\tilde{\mathbf{v}}_2^{(0)},...,\tilde{\mathbf{v}}_{N}^{(0)}\}$, where each particle represents a possible solution for the APV, i.e.,
\begin{equation}
\label{particle}
\begin{aligned}
\tilde{\mathbf{r}}_{n}^{(0)}=[\underbrace{x_{n,1}^{(0)},y_{n,1}^{(0)}}_{\text{MA}~~1},\underbrace{x_{n,2}^{(0)},y_{n,2}^{(0)}}_{\text{MA}~~2},\cdots,\underbrace{x_{n,M}^{(0)},y_{n,M}^{(0)}}_{\text{MA}~~M}]^T,
\end{aligned}
\end{equation}
where $x_{n,m}^{(0)},y_{n,m}^{(0)}\sim \mathcal{U}[-A/2,A/2]$ for $1\le n \le N, 1\le m \le M$
ensures that the initial position of each MA does not exceed the finite moving region, i.e., constraint~(\ref{op_APVB}) holds.

Then, each particle updates its position according to the individual experience (the known local best position, i.e., $\tilde{\mathbf{r}}_{n,pbest}$) and the swarm experience (the known global best position, i.e., $\tilde{\mathbf{r}}_{gbest}$).
Thus, for each iteration, the velocity and position of each particle are updated as
\begin{equation}
	\label{update_velocity}
	\tilde{\mathbf{v}}_{n}^{(t+1)}=\omega \tilde{\mathbf{v}}_{n}^{(t)} + c_1 \tau_1 \left(\tilde{\mathbf{r}}_{n,pbest}-\tilde{\mathbf{r}}_{n}^{(t)}\right)+c_2 \tau_2 \left(\tilde{\mathbf{r}}_{gbest}-\tilde{\mathbf{r}}_{n}^{(t)}\right),
\end{equation}
\begin{equation}
	\label{update_position}
	\tilde{\mathbf{r}}_{n}^{(t+1)}=\mathcal{B}\left( \tilde{\mathbf{r}}_{n}^{(t)} + \tilde{\mathbf{v}}_{n}^{(t+1)} \right),
\end{equation}
for $1 \le n \le N$, with $t$ representing the iteration index. Parameters $c_1$ and $c_2$ are the individual and global learning factors, which represent the step size of each particle moving toward the best position.  $\tau_1$ and $\tau_2$ are two random parameters uniformly distributed in $[0,1]$,  which aim to increase the randomness of the search  for escaping from local optima. $\omega$ is the inertia weight, which is used to maintain the inertia of the particle movement. In particular, in order to balance the speed and accuracy of particle swarm search, the inertia weight is continuously decreased during the iterations as follows:
\begin{equation}
	\label{inertia}
	\omega=\left(\omega_{\mathrm{max}}-\dfrac{(\omega_\mathrm{max}-\omega_\mathrm{min})t}{T}\right),
\end{equation}
where $\omega_{\mathrm{max}}$ and $\omega_{\mathrm{min}}$ are the minimum value and maximum value of $\omega$ and $T$ is the maximum iteration number. 

Due to constraint (\ref{op_APVB}), if a particle moves out of the boundary of the feasible region, we project its position component to the corresponding minimum/maximum value, i.e., 
\begin{equation}\label{bound_p}
	[\mathcal{B}(\tilde{\mathbf{r}})]_i=\left\{
	\begin{aligned}
		&-\frac{A}{2}, ~\text{if } [\tilde{\mathbf{r}}]_i<-\frac{A}{2},\\
		&\frac{A}{2},~~~~\text{if }[\tilde{\mathbf{r}}]_i>\frac{A}{2},\\
		&\left[\tilde{\mathbf{r}}\right]_{i},~~\text{otherwise}.
	\end{aligned}\right.
\end{equation}
The utilization of projection function $\mathcal{B}(\tilde{\mathbf{r}})$ in~(\ref{update_position}) is to ensure that the solution for APV is always located in the feasible region during the iterations.

The fitness of each particle is evaluated according to Algorithm~\ref{BCD} and is given by $R(\tilde{\mathbf{r}}_{n}),1 \leq n \leq N$, for maximizing the minimum achievable rate of multiple users under the given APV. Moreover, in order to ensure constraint (\ref{op_APVC}), we introduce an adaptive penalty factor ~\cite{DE2011DS} to the fitness function and update it as follows:
\begin{equation}
\label{penalty}
\mathcal{F}\left(\tilde{\mathbf{r}}_{n}^{\left(t\right)}\right)=R\left(\tilde{\mathbf{r}}_{n}^{\left(t\right)}\right)-\tau \left|\mathcal{P}\left(\tilde{\mathbf{r}}_{n}^{\left(t\right)}\right)\right|,
\end{equation}
where $\mathcal{P}\left(\tilde{\mathbf{r}}\right)$ is a set, in which each element represents a pair positions of MAs in the APV $\tilde{\mathbf{r}}$ that violate the minimum inter-MA distance constraint. It can be defined as
\begin{equation}
	\label{powerset}
	\mathcal{P}\left(\tilde{\mathbf{r}}\right)=\left\{ (\mathbf{r}_{m},\mathbf{r}_{i})| \left\|\mathbf{r}_{m}-\mathbf{r}_{i}\right\|_{2} < D, 1 \le  m < i \le M\right\}.
\end{equation}
$\tau$ is a large positive penalty parameter which ensures that the inequality equation  $R\left(\tilde{\mathbf{r}}_{n}^{\left(t\right)}\right)-\tau \le 0$ holds for all APVs. 
Thus, the penalty parameter drives each particle to move to positions in which the minimum inter-MA distance is guaranteed, since otherwise its fitness value falls to less than zero. It means that during the iterations,  $\left|\mathcal{P}\left(\tilde{\mathbf{r}}_{n}^{\left(t\right)}\right)\right|$ will approach zero, i.e., constraint (\ref{op_APVC}) is satisfied eventually.

With the fitness evaluation conducted on each particle, their individual and global best positions are updated until convergence. The final best position among the particles is generally a suboptimal solution for APV, and its corresponding  receive combining matrix and transmit power matrix are calculated by Algorithm \ref{BCD}.

The detailed PSO-based overall algorithm for solving problem (\ref{op}) is summarized in Algorithm~\ref{PSO}.
In line 1, in the $2M$-dimensional search space, the position and velocity of each particle are randomly initialized with each component uniformly distributed in $[-A/2,A/2]$. In lines 2-3, each particle is evaluated by the fitness function, thus finding the local and global best positions. In line 5, the inertia weight linearly decreases from $\omega_{\mathrm{max}}$ to $\omega_{\mathrm{min}}$ with the increasing iteration index. In line 7, the velocity of each particle is updated according to the relative local and global best positions, which drive the particle moving in the feasible region. In lines 8-14, we evaluate the particle’s fitness value and compare it with that of its local/global best position. 
For each particle, if its fitness value is better than that of its local best position or the global best position, then the corresponding best locations are replaced with the current particle’s position. 
Thus, in lines 4-16, the global best position can be updated with its fitness value non-decreasing during the iterations.
Hereto, we have solved the original problem~(\ref{op}). In the proposed solution,
the receive combining matrix and transmit power matrix are optimal,
while the APV is suboptimal in general.
\begin{algorithm}[t] \small
    \label{PSO}
    \DontPrintSemicolon
    \SetAlgoLined
    \caption{PSO-based Algorithm for solving problem~(\ref{op}).}
    \begin{algorithmic}[1]
        \REQUIRE ~$M$, $K$, $\mathcal{C}_r$, $\lambda$, $p_{\mathrm{max}}$, $\sigma^2$, $\{\mathbf g_k\}$, $\{\theta_{k,l}\}$, $\{\phi_{k,l}\}$, $N$, $T$,\\ $\epsilon$, $\xi$, $c_1$, $c_2$, $\omega_{\mathrm{min}}$, $\omega_{\mathrm{max}}$, $\tau$.
        \ENSURE ~$\tilde{\mathbf r},\mathbf{W}, \mathbf{P}$. \\
        \STATE Initialize the $N$ particles with positions $\mathcal{R}^{(0)}$ and velocities $\mathcal{V}^{(0)}$.
        \STATE Evaluate the fitness value for each particle using Algorithm 2.
        \STATE Obtain the local best position $\tilde{\mathbf{r}}_{n,pbsest}=\tilde{\mathbf{r}}_{n}^{\left(0\right)}$ for $1\le n \le N$ and the global best position $\tilde{\mathbf{r}}_{gbest}=\mathop{\arg\max}\limits _{\tilde{\mathbf{r}}_{n}^{\left(0\right)}}\{\mathcal{F}\left(\tilde{\mathbf{r}}_{1}^{\left(0\right)}\right),\mathcal{F}\left(\tilde{\mathbf{r}}_{2}^{\left(0\right)}\right),...,\mathcal{F}\left(\tilde{\mathbf{r}}_{N}^{\left(0\right)}\right)\}$.
        \FOR{$t=1$ to $T$}
            \STATE Calculate the inertia weight $\omega$ according to (\ref{inertia}).
            \FOR{$n=1$ to $N$}
                \STATE Update the velocity and position of particle $n$  according to (\ref{update_velocity}) and (\ref{update_position}), respectively.
                \STATE  Evaluate the fitness value for particle $n$ using Algorithm \ref{BCD} and update it according to~(\ref{penalty}), i.e., $\mathcal{F}\left(\tilde{\mathbf{r}}_{n}^{\left(t\right)}\right)$.
                \IF{$\mathcal{F}\left(\tilde{\mathbf{r}}_{n}^{\left(t\right)}\right)>\mathcal{F}\left (\tilde{\mathbf{r}}_{n,pbest}\right)$}
                \STATE Update $\tilde{\mathbf{r}}_{n,pbest}\leftarrow \tilde{\mathbf{r}}_{n}^{(t)}$.
                \ENDIF
                \IF{$\mathcal{F}\left(\tilde{\mathbf{r}}_{n}^{\left(t\right)}\right)>\mathcal{F}\left (\tilde{\mathbf{r}}_{gbest}\right)$}
                \STATE Update $\tilde{\mathbf{r}}_{gbest}\leftarrow \tilde{\mathbf{r}}_{n}^{(t)}$.
                \ENDIF
            \ENDFOR
        \ENDFOR
        \STATE Obtain the suboptimal APV $\tilde{\mathbf{r}}=\tilde{\mathbf{r}}_{gbest}$.
        \STATE Calculate the corresponding receive combining matrix $\mathbf{W}$ and transmit power matrix $\mathbf{P}$ according to Algorithm~\ref{BCD}. 
        \RETURN $\tilde{\mathbf r},\mathbf{W}, \mathbf{P}$.
    \end{algorithmic}
\end{algorithm}

\subsection{Convergence and Complexity Analysis}\label{sec_ConvergenceComplexity}
Since the overall algorithm is two-loop based, its convergence depends on the convergence of BCD-based algorithm in the inner-loop and PSO-based algorithm in the outer-loop. 
The convergence of  Algorithm \ref{BCD} is guaranteed by the following inequality:
\begin{equation}
	\label{convergence_inner}
	\begin{split}
		\mathcal{G}\left(\mathbf{P}^{(j)},\mathbf W^{(j)}\right)
		&=
		\mathcal{G}\left(\mathbf{P}^{(j)},\hat{\mathbf W}\left(\tilde{\mathbf r},\mathbf P^{(j)}\right)\right)\\
		&\overset{(a)}{\ge} 
		\mathcal{G}\left(\mathbf{P}^{(j)},\hat{\mathbf W}\left(\tilde{\mathbf r},\mathbf P^{(j-1)}\right)\right)\\
		&\overset{(b)}{\ge} 
		\mathcal{G}\left(\mathbf{P}^{(j-1)},\hat{\mathbf W}\left(\tilde{\mathbf r},\mathbf P^{(j-1)}\right)\right)\\
		&=
		\mathcal{G}\left(\mathbf{P}^{(j-1)},\mathbf W^{(j-1)}\right),
	\end{split}
\end{equation}
where $(a)$ holds because $\hat{\mathbf W}\left(\tilde{\mathbf r},\mathbf P^{(j)}\right)$ is the optimal MMSE combining matrix for maximizing the SINR of each user under the current transmit power $\mathbf{P}^{(j)}$, and $(b)$ holds since $\mathbf{P}^{(j)}$ is the optimal transmit power searched by bisection method under the current MMSE combining matrix $\hat{\mathbf W}\left(\tilde{\mathbf r},\mathbf P^{(j-1)}\right)$. It means that the objective value is non-decreasing during the iterations in Algorithm~\ref{BCD}.

Moreover, the fitness value of the global best position is non-decreasing during the iterations in Algorithm~\ref{PSO}, i.e.,
\begin{equation}
	\label{convergence_outter}	
	\mathcal{F}\left(\tilde{\mathbf r}_{gbest}^{\left(t+1\right)}\right)\ge \mathcal{F}\left(\tilde{\mathbf r}_{gbest}^{\left(t\right)}\right).
\end{equation} 
Meanwhile, the objective value of problem (\ref{op}) is always bounded. Thus, the convergence of the overall algorithm is guaranteed. Moreover, the convergence performance will also be validated by simulation in Section~~\ref{sec_SimulationResults}.

The computational complexity of Algorithm~\ref{bisection} for solving problem (\ref{op_power}) is $\mathcal{O}(K^3\log_{2}{\epsilon^{-1}})$, depending on the search accuracy $\epsilon$ and the number of users $K$.
Denoting the maximum number of iterations of Algorithm~\ref{BCD} for 
solving problem (\ref{op:PandW}) as $J$, the corresponding computational complexity is given by $\mathcal{O}\left(J(M^3+K^3\log_{2}{\epsilon^{-1}} )\right)$.
As a result, with the swarm size $N$ and the maximum number of iterations $T$, the maximum computational complexity of Algorithm~\ref{PSO} for solving problem (\ref{op}) is $\mathcal{O}\left(NTJ(M^3+K^3\log_{2}{\epsilon^{-1}} )\right)$.

\section{Simulation Results}\label{sec_SimulationResults}
In this section, numerical simulation results are presented to evaluate the performance of our proposed MA-aided multiuser communication system and demonstrate the effectiveness of our proposed algorithms for maximizing the minimum achievable rate among users.
\subsection{Simulation Setup}
In the simulations, we consider a scenario where $K$ FPA-users are served by the BS equipped with $M$ MAs, and the distance between user $k$ and the BS is
assumed to be a random variable following uniform distributions, i.e., $d_k \sim \mathcal{U}[20,100], 1\le k \le K$. The moving region for MAs at the BS is set as a square area of size $[-A/2,A/2]\times [-A/2,A/2]$. We adopt the geometry channel model, in which the numbers of receive paths for all users are the same, i.e., $L_k=L, 1\le k \le K$. For each user, each element of the PRV is an i.i.d. CSCG random varible, i.e., $g_{k,l} \sim \mathcal{CN}(0,\rho d_k^{-\alpha}/L), 1\le k \le K, 1 \le l \le L$, where $\rho d_k^{-\alpha}$ is the expected channel gain of user $k$ with $\rho$ representing the path loss at the reference distance of 1 meter (m) and $\alpha$ denoting the path loss exponent. Dividing the expected channel gain by the number of paths is aimed to guarantee the same average channel gain under different numbers of paths. The elevation and azimuth AoAs for each user are assumed to be i.i.d. variables following the uniform distribution over $[-\pi/2,\pi/2]$, i.e., $\theta_{k,l},\phi_{k,l} \sim \mathcal{U}[-\pi/2,\pi/2], 1\le k \le K, 1 \le l \le L$. The detailed settings of simulation parameters are listed in Table~\ref{tab:para}, unless specified otherwise. Each point in the simulation figures is the average result over $10^3$ random user distributions and channel realizations.
\begin{table}[h]
	\caption{Simulation Parameters}\label{tab:para}
	\footnotesize
	\begin{center}
		\begin{tabular}{|c|l|c|}
			\hline
			\textbf{Parameter}                        &\multicolumn{1}{c|}{\textbf{Description}}                                       & \textbf{Value} \\
			\hline
			$M$                                 & Number of MAs at the BS                                  	& 16 \\
			\hline
			$K$                                 & Number of users                                   & 12 \\
			\hline
			$L$                                 &Number of channel paths for each user                           & 10 \\
			\hline
			$\lambda$                       	&Carrier wavelength                           	& 0.1 m \\
			\hline
			$A$                       &Length of the sides of receive region                                  & $3\lambda$ \\
			\hline
			$D$                      & Minimum inter-MA distance                                      & $\lambda/2$ \\
			\hline
			$\rho$                      & Channel gain at the reference distance                                & $-40$ dB \\
			\hline
			$\alpha$                    & Path loss exponent                          		& 2.8 \\
			\hline
			$\sigma^2$                    & Average noise power                           		& $-80$ dBm \\
			\hline
			$p_{\mathrm{max}}$                    & Maximum transmit power for each user                          		& 10 dBm \\
			\hline
			$\epsilon$                    & Convergence threshold in Algorithm~\ref{bisection}   			& $10^{-3}$ \\
			\hline
			$\xi$                    & Convergence threshold in Algorithm~\ref{BCD}   					& $10^{-3}$ \\
			\hline
			$N$                    & Number of particles in Algorithm~\ref{PSO}                 & 200 \\
			\hline
			$T$                    & Maximum number of iterations in Algorithm~\ref{PSO}              & 300 \\
			\hline
			$c_1$                    & Individual learning factor in Algorithm~\ref{PSO}                 & 1.4 \\
			\hline
			$c_2$                    & Global learning factor in Algorithm~\ref{PSO}              & 1.4 \\
			\hline
			$\omega_{\mathrm{min}}$                    & Minimum inertial weight in Algorithm~\ref{PSO}                 & 0.4 \\
			\hline
			$\omega_{\mathrm{max}}$                    & Maximum inertial weight in Algorithm~\ref{PSO}              & 0.9 \\
			\hline
			$\tau$                    & Penalty parameter in Algorithm~\ref{PSO}              & 10 \\
			\hline				
		\end{tabular}
	\end{center}
\end{table}
\subsection{Convergence Performance of Proposed Algorithms}

\begin{figure}[t]
	\begin{center}
		\includegraphics[width=\figwidth cm]{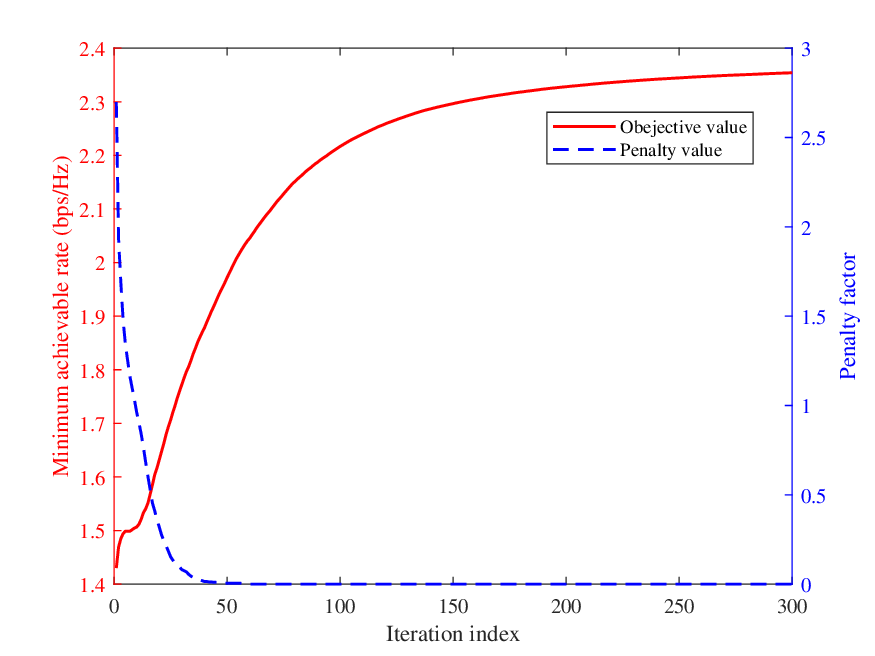}
		\caption{{\color{black}Objective value and penalty value versus iteration index for the proposed algorithm.}}
		\label{fig:OvsP}
	\end{center}
\end{figure}

\begin{figure}[t]
	\begin{center}
		\includegraphics[width=\figwidth cm]{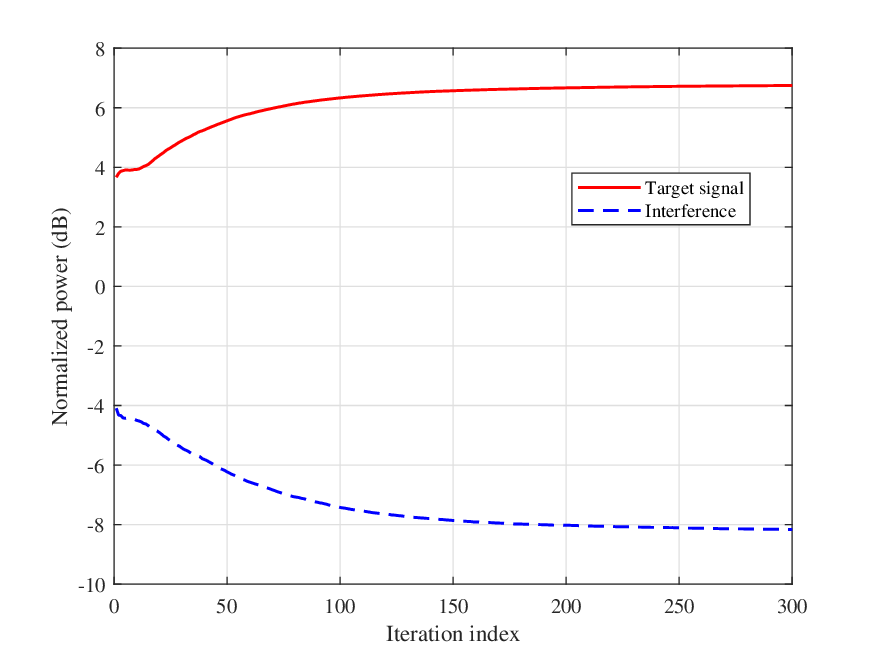}
		\caption{{\color{black}Average receive powers of the target signals and
				interference for multiple users normalized by noise power
				versus iteration index.}}
		\label{fig:TvsI}
	\end{center}
\end{figure}

In Fig.~\ref{fig:OvsP}, the convergence of the proposed algorithms for the MA-aided multiuser communication system is presented. Moreover, in order to validate the effectiveness of our proposed adaptive penalty factor in~(\ref{penalty}), we illustrate the penalty value versus the iteration index.
As can be observed, the minimum achievable rate of all users increases with the iteration index and remains nearly unchanged after 250 iterations, which demonstrates fast convergence performance.
In addition, the penalty value remains zero after 50 iterations, which guarantees that the minimum inter-MA distance constraint is satisfied. Particularly, the minimum achievable rate of all users increases from 1.44 bits per second per hertz (bps/Hz) to 2.36 bps/Hz, which yields about $63\%$ performance improvement.

In Fig.~\ref{fig:TvsI}, we show the convergence performance of normalized receive powers of the target signals and interference, where the powers of the target signal and interference for each user are normalized by noise power, i.e., the average values of $p_k[\mathbf{A}]_{k,k}/b_k$ and $\sum_{i=1,i\ne k}^{K}p_i[\mathbf{A}]_{k,i}/b_k$ for $1\le k \neq i \le K$. During the iterations, on one hand, it can be observed that the normalized signal power increases from $4\text{ dB}$ to $6.7\text{ dB}$. It indicates that APV optimization can leverage the spatial DoFs, which is beneficial for obtaining higher channel power gain. On the other hand, the normalized interference power decreases from $-4\text{ dB}$ to $-8.1\text{ dB}$, which validates that the proposed algorithm can decrease the correlation of the channel vectors and mitigate interference among multiple users. The results reveal that the proposed algorithm for the MA-aided multiuser communication system can effectively enhance rate performance via channel gain improvement and multiuser interference suppression. 
\subsection{Channel Characteristics under APV Optimization}
\begin{figure*}[t]
	\centering
	\subfigure[]{
		\includegraphics[width=\figwidth cm]{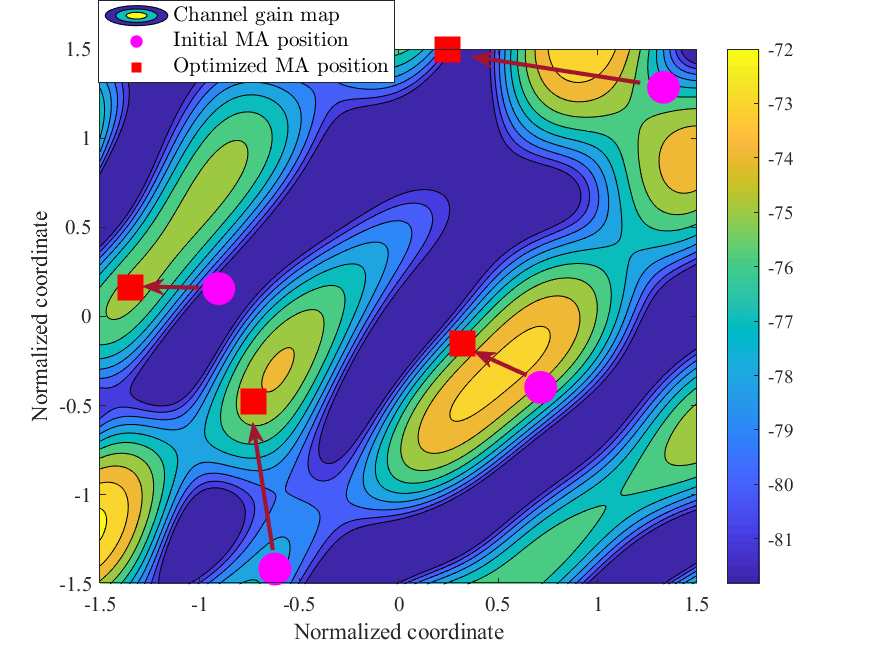}
	} 
	\hspace{10mm} 
	\subfigure[]{
		\includegraphics[width=\figwidth cm]{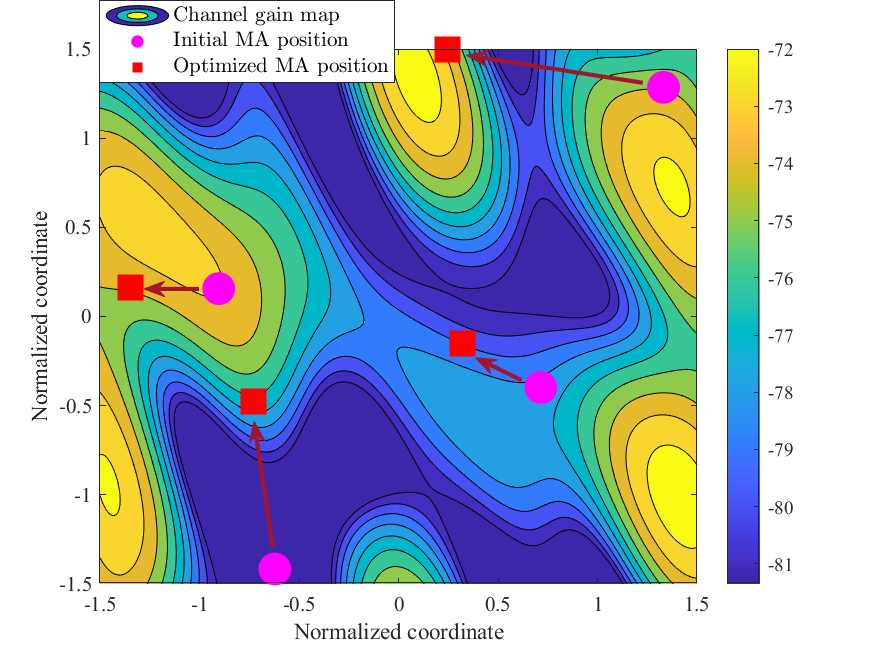}
	}
	\caption{Channel power gain (dB) between MAs of the BS and each single FPA of two users: (a) User 1; (b) User 2.}
	\label{channel_map}
\end{figure*}
To validate the impact of APV optimization on improving the channel conditions, one realization of the channel power gain (in dB) versus the MAs' positions
for the two-user case is illustrated in Fig.~\ref{channel_map}. For the sake of visualization, we set 4 MAs deployed at the BS. It is shown that each user has its unique channel gain map in the receive region. The APV optimization should consider the trade-off between the channel gain maps of two users and tends to choose the positions with more favorable channel conditions for both users. In order to numerically verify our proposed  algorithms, we define 
the channel power gain and the normalized channel cross-correlation as $||\mathbf{h}_k(\tilde{\mathbf{r}})||_2^2$ and $\frac{|\mathbf{h}_k(\tilde{\mathbf{r}})^{\mathrm{H}}\mathbf{h}_i(\tilde{\mathbf{r}})|}{||\mathbf{h}_k(\tilde{\mathbf{r}})||_2^2||\mathbf{h}_i(\tilde{\mathbf{r}})||_2^2}$ for $1\le k \neq i \le K$, respectively. 
By employing the proposed
algorithms, the channel power gains of user 1 and user 2 increase from $-$70.1 dB and $-$73.1 dB to $-$67.2 dB and $-$67.2 dB, respectively. It indicates that the MAs move from the initial positions with low channel power gains to the optimized positions with high channel power gains. Furthermore, APV optimization tends to enable two users with approximately equal channel gains to guarantee their max-min fairness. In addition, the normalized channel cross-correlation between user 1 and user 2 decreases from 0.1461 to 0.0518. The channel correlation is significantly reduced by APV optimization, thus suppressing multiuser interference.
\subsection{Performance Comparison with Benchmark Schemes}
The proposed algorithm is labeled by ``\textbf{MA}". {\color{black}Besides, three benchmark schemes are defined for performance comparison, namely ``\textbf{FPA}", ``\textbf{APS}'' and ``\textbf{MPZF}", respectively. }
\begin{itemize}
	\item {\textbf{FPA}: The BS is equipped with FPA-based uniform planar array with $M$ antennas, spaced by $\lambda/2$. The corresponding optimal receive combining matrix and transmit power matrix are obtained by Algorithm~\ref{BCD}. }

	\item {\textbf{APS}: The receive moving region is quantized into discrete locations with equal-distance $\lambda/2$. Since the exhaustive search over all possible discrete locations for each MA leads to prohibitive computational complexity, we adopt the alternating position selection (APS) method to alternately select each MA's position with the others being fixed. It means that only one MA's position is optimized while other MAs' positions are fixed for each iteration. Similarly, Algorithm~\ref{BCD} is performed to calculate receive combining and transmit power matrices. 
	}
	
	\item {\textbf{MPZF}: This scheme performs the similar APV optimization method with the MA scheme. However, the receive combining matrix is given by a zero-forcing (ZF) receiver instead of an MMSE receiver, and the corresponding optimal transmit power control strategy is that each user transmits with the maximum power, i.e., maximum-power zero-forcing (MPZF).}
	
\end{itemize}

\begin{figure}[t]
	\begin{center}
		\includegraphics[width=\figwidth cm]{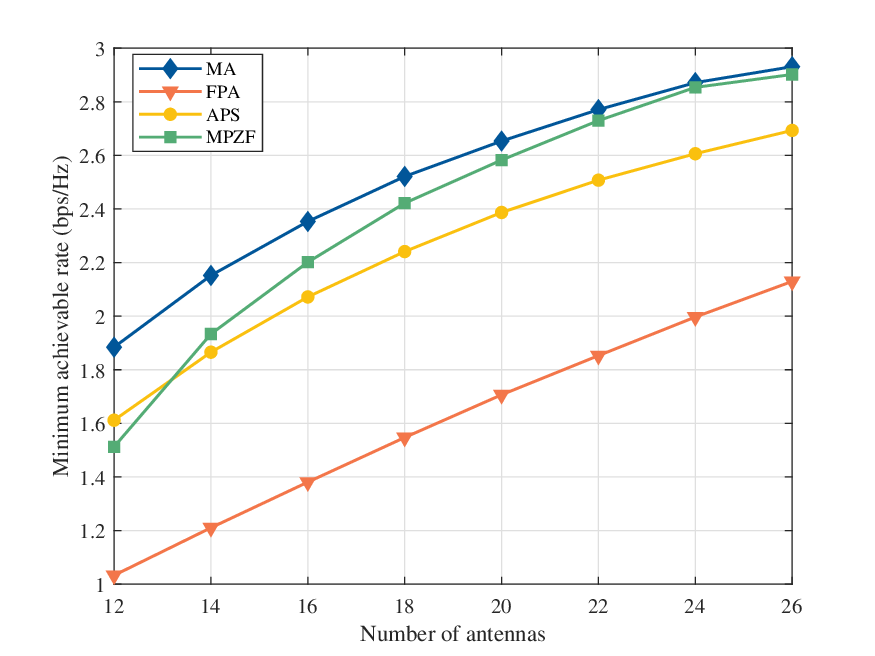}
		\caption{{\color{black}Minimum achievable rates for different schemes versus number of antennas.}}
		\label{fig:RvsM}
	\end{center}
\end{figure}

\begin{figure}[t]
	\begin{center}
		\includegraphics[width=\figwidth cm]{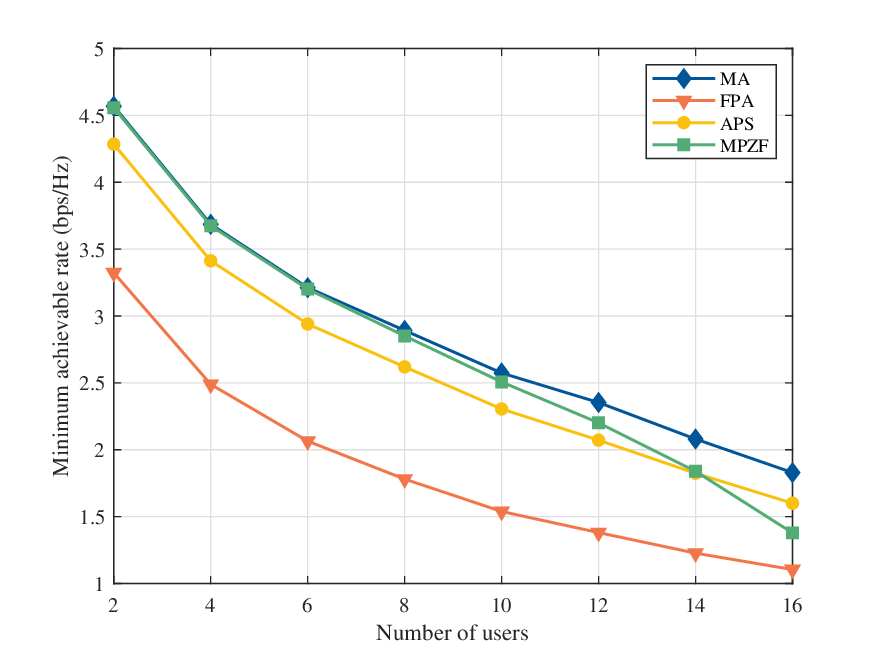}
		\caption{{\color{black}Minimum achievable rates for different schemes versus number of users.}}
		\label{fig:RvsK}
	\end{center}
\end{figure}

\begin{figure}[t]
	\begin{center}
		\includegraphics[width=\figwidth cm]{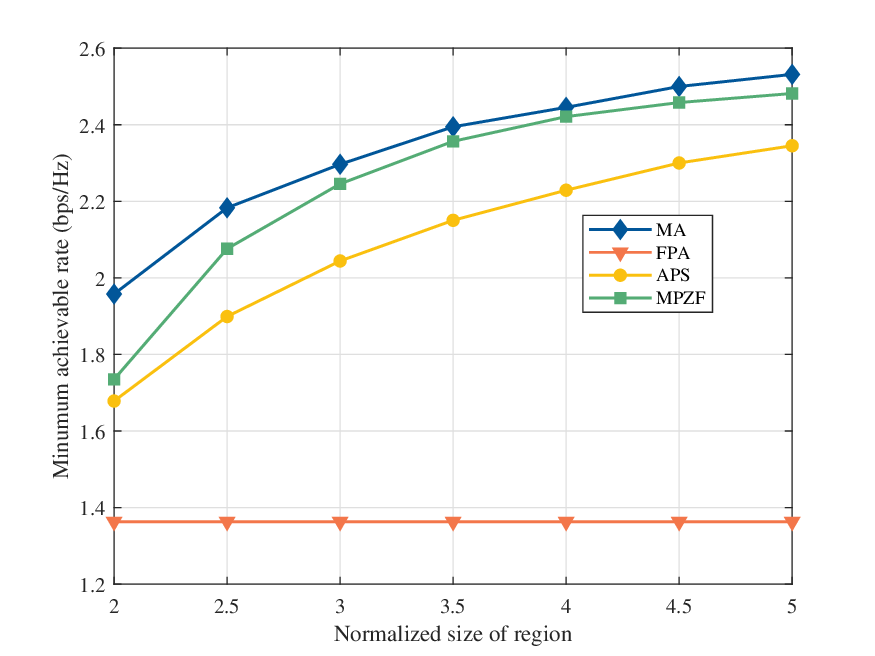}
		\caption{{\color{black}Minimum achievable rates for different schemes versus normalized size of region.}}
		\label{fig:RvsA}
	\end{center}
\end{figure}

\begin{figure}[t]
	\begin{center}
		\includegraphics[width=\figwidth cm]{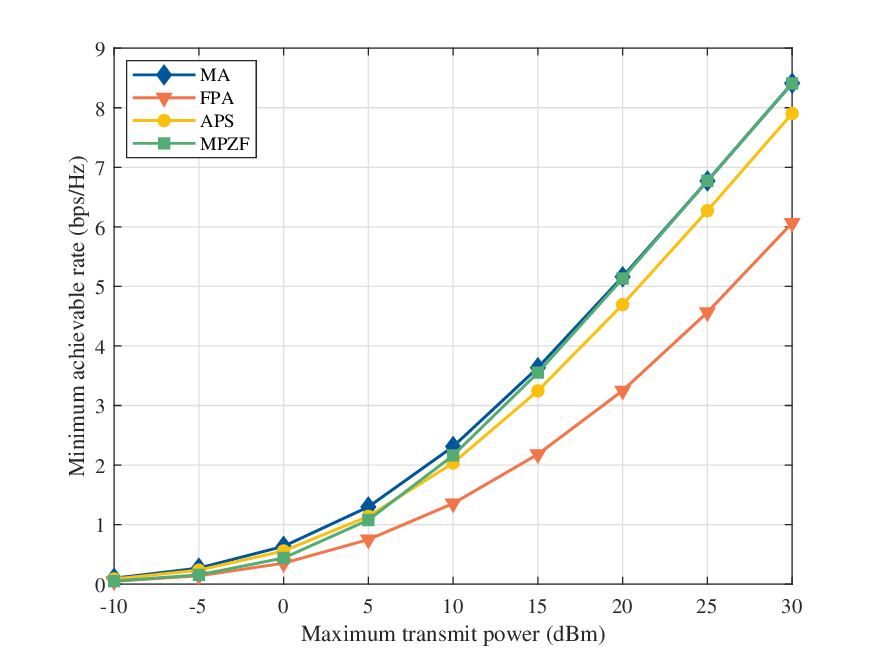}
		\caption{{\color{black}Minimum achievable rates for different schemes versus maximum transmit power.}}
		\label{fig:RvsP}
	\end{center}
\end{figure}

\begin{figure}[t]
	\begin{center}
		\includegraphics[width=\figwidth cm]{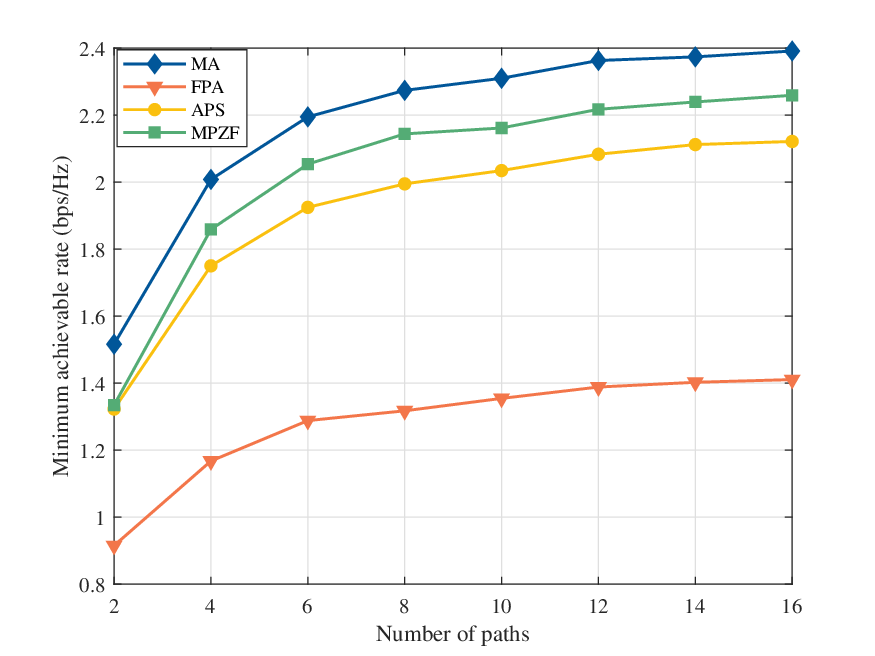}
		\caption{{\color{black}Minimum achievable rates for different schemes versus number of paths.}}
		\label{fig:RvsL}
	\end{center}
\end{figure}

In Fig~\ref{fig:RvsM}. we compare the minimum achievable rates for different schemes versus the number of antennas. It can be seen that the proposed scheme outperforms all other benchmark schemes. With the increase of antennas, the minimum achievable rate increases because there is an improvement in spatial diversity gain  and beamforming gain. Compared to the conventional FPA and APS schemes, all MAs' positions in the MA scheme are jointly optimized in the continuous spatial field, thus leveraging spatial DoF to significantly reduce the number of antennas required for the same rate performance. Besides, the performance gap between MA and MPZF schemes decreases because the channel correlation among multiple users becomes smaller. 

Fig.~\ref{fig:RvsK} compares the minimum achievable rates for different schemes versus the number of users. As can be observed, the proposed scheme still outperforms
all other benchmark schemes. In addition, the minimum achievable rates decrease with the number of users for all schemes. The reason is that the increasing number of users leads to more severe mutual interference. Besides, as the number of users increases, the performance gap between the MPZF and MA schemes increases, and the performance of the MPZF scheme degrades to be worse than that of the APS scheme for $K=16$. It indicates that MMSE-based receive combining matrix and the corresponding transmit power matrix can significantly mitigate multiuser interference. 
With the conventional FPAs, the FPA scheme behaves the worst performance for all settings because the wireless channels' spatial DoFs are not properly exploited to obtain higher interference mitigation gain and channel power gain.

Fig.~\ref{fig:RvsA} compares the minimum achievable rates for different schemes versus the normalized region size for moving antennas at the BS, where the size of the moving region is normalized
by carrier wavelength, i.e., $A/\lambda$. As can be observed, with the increasing normalized region size, the minimum achievable rate achieved by the MA scheme increases with a decreasing speed. This is because the increase of moving region size can further explore the DoFs in the spatial domain, such that MAs can be deployed at positions with higher channel power gains and lower multiuser interference. 
In particular, the decreasing speed indicates that it is not required to move the antennas in a large region if the number of channel paths is limited.

Fig.~\ref{fig:RvsP} compares the minimum achievable rates for different schemes versus the maximum transmit power of each user, which impacts the received signal-to-noise ratio (SNR). The minimum achievable rate significantly increases with the maximum transmit power for all schemes and the MA scheme outperforms all other schemes in terms of minimum achievable rate. In the low-SNR regime, such as $p_{\mathrm{max}}=0\text{ dBm}$, the MA scheme has a significant advantage over the MPZF scheme, while this advantage almost disappears in the high-SNR regime, such as $p_{\mathrm{max}}=30\text{ dBm}$. This is because compared to MMSE receivers, ZF receivers yield amplified noise power in the received signal, which is influential to the performance in the low-SNR regime and negligible in the high-SNR regime.

In Fig.~\ref{fig:RvsL}, we compare the minimum achievable rates for different schemes versus the number of channel paths. The minimum achievable rate increases with the number of channel paths for all schemes since more paths lead to higher multi-path diversity gain and lower correlation among the channel vectors for multiple users. Besides, the MA scheme can make use of the channel variation to further reduce the channel correlation and have better spatial diversity. Thus, the increasing rate of the MA scheme is much higher than that of the FPA scheme. In addition, it is shown that the minimum achievable rate stays at a very large value and is almost unchanged when the number of paths is large enough, i.e., 14 or 16. This is because a larger region is required for antenna moving to fully exploit the wireless channel spatial variation under increasing number of channel paths. 
\subsection{Impact of Imperfect FRI}
\begin{figure}[t]
	\begin{center}
		\includegraphics[width=\figwidth cm]{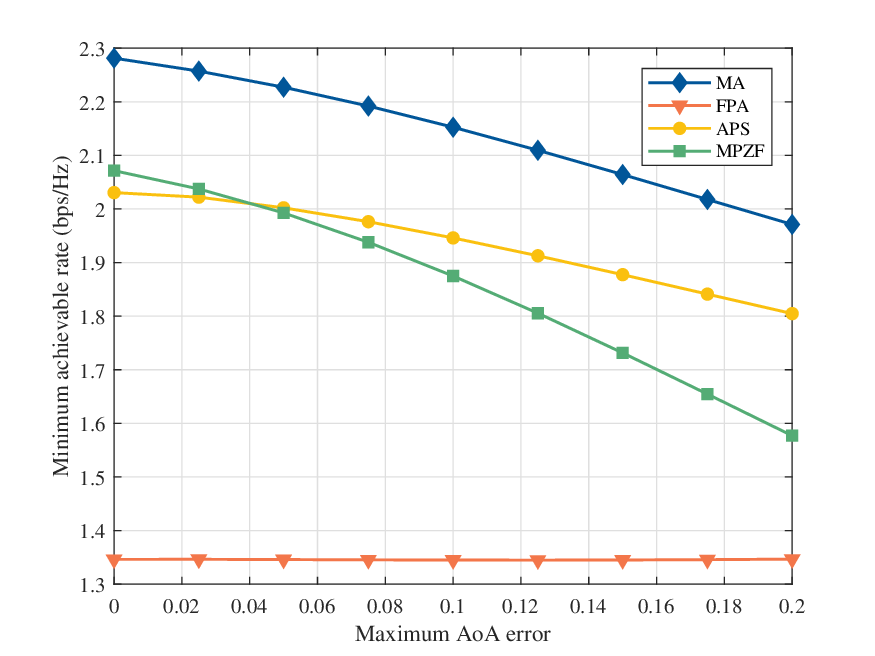}
		\caption{{\color{black}Impact of the AoA error on the performance of the
				proposed algorithms.}}
		\label{fig:AoA_error}
	\end{center}
\end{figure}
\begin{figure}[t]
	\begin{center}
		\includegraphics[width=\figwidth cm]{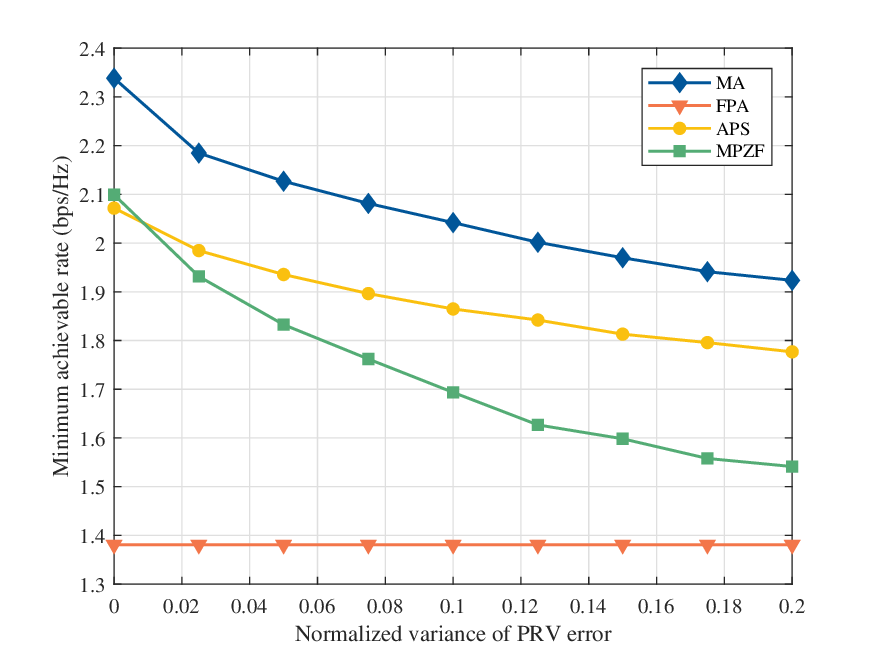}
		\caption{{\color{black}Impact of the PRV error on the performance of the
				proposed algorithms.}}
		\label{fig:PRV_error}
	\end{center}
\end{figure}
It is worth noting that the aforementioned solutions and simulation results
are based on the assumption of perfect FRI
(including AoAs and PRVs) at the BS. Nevertheless, 
though there are some possible solutions in channel estimation for MA-aided communication
systems~\cite{ma2023compressed}, the FRI errors are generally inevitable in practice, 
which results in performance degradation. Thus, it is necessary to take FRI errors into account in our considered systems and evaluate their impact on communication performance. For this purpose, we introduce two factors to characterize FRI errors. On one hand,
let $\hat\theta_{k,l}$ and $\hat \phi_{k,l}$ denote the estimated AoAs of the $l$-th channel path between the BS and user
$k$, $1\le k \le K,1\le l \le L$. We define the AoA errors as the differences between actual AoAs and estimated AoAs, which are assumed to be i.i.d. uniformly distributed, i.e., $\theta_{k,l}- \hat\theta_{k,l}\sim \mathcal{U}[-\mu/2, \mu/2]$ and $\phi_{k,l}- \hat\phi_{k,l}\sim \mathcal{U}[-\mu/2, \mu/2]$ with $\mu$ representing the maximum AoA error. On the other hand, let $\hat g_{k,l}$ denote the estimated path-response coefficient of the $l$-th channel path between the BS and user
$k$, $1\le k \le K,1\le l \le L$. The PRV errors are defined as the differences between actual PRVs and estimated PRVs, which are assumed to be i.i.d. CSCG random variables, i.e., $\frac{g_{k,l}-\hat g_{k,l}}{|g_{k,l}|}\sim \mathcal{CN}(0, \delta)$ with $\delta$ denoting the variance of the normalized PRV error.

In Fig.~\ref{fig:AoA_error}, we show the impact of the AoA error on the performance of the proposed algorithm, where the PRVs between the BS and users are assumed to be accurate for optimization. In order to focus on studying the effect of imperfect FRI on APV optimization, the APV optimization is solved by Algorithm~\ref{PSO} based on the estimated AoAs. Then, for given suboptimal APV, the receive combining matrix and transmit power matrix are calculated by Algorithm~\ref{BCD} based on the actual AoAs, thus achieving the max-min achievable rate of multiple users. From Fig.~\ref{fig:AoA_error}, we can find that the MA scheme suffers achievable rate performance degradation as the AoA error increases. However, though the AoA error is large (such as $\mu=0.2$), the MA scheme still significantly outperforms the FPA scheme in terms of the minimum achievable rate among users. Besides, the performance of the MPZF scheme is degraded to be worse than that of the APS scheme, and the performance gap between two schemes becomes larger with the increase of the normalized variance of PRV. This is because a larger AoA error may lead to MAs being deployed in positions with higher multiuser interference, which is detrimental to the performance of the ZF receiver. 

Finally, the impact of the PRV error on the performance of the proposed algorithm is presented in Fig.~\ref{fig:PRV_error}, where the AoAs of the channel paths between the BS and users are assumed to be accurate for optimization. Similarly, the suboptimal APV is given based on the estimated PRVs while the receive combining matrix and transmit power matrix are calculated based on the actual PRVs. From Fig.~\ref{fig:PRV_error}, it can be seen that the MA scheme achieves better performance over the benchmark schemes even for a high PRV error, whereas the improvement is diminishing as the normalized variance of PRV error increases due to the inaccurate PRVs for APV optimization. Similarly, the performance of the MPZF scheme is degraded due to the multiuser interference caused by PRV errors.
\section{Conclusion}\label{sec_Conclusion}
In this paper, we proposed a new BS architecture with multiple MAs deployed to improve the multiuser communication rate performance as compared to traditional BSs mounted with FPAs. We first modeled the multiuser channel as a function of the APV to characterize the multi-path response between the multiple MAs at the BS and the single FPA at each user. Then, based on this channel model, a joint optimization problem was formulated for designing the MAs positioning, receive combining, and transmit power control for each user to maximize the minimum achievable rate among multiple users, under the constraints of finite moving region for MAs, minimum inter-MA distance, and maximum transmit power of each user. To solve this non-convex optimization problem efficiently, we developed a two-loop iterative algorithm based on PSO. 
Simulation results demonstrated that compared to FPA-based systems, our proposed solution for MA-aided uplink multiuser communication systems can significantly improve the rate performance by exploiting the new design DoF via antenna position optimization. Moreover, we evaluated the impact of imperfect FRI on the optimized solution for APV. The results revealed that the proposed solution can achieve robust performance against the estimation errors of the AoAs and PRVs in the FRI-based multiuser channel.


\appendices
\bibliographystyle{IEEEtran}
\bibliography{IEEEabrv,MA}
\end{document}